 \documentclass[final,5p,times,twocolumn]{elsarticle}

\usepackage{amssymb}
\usepackage{amsmath}
\usepackage{amsthm}
\usepackage{amsfonts}

\newcommand{\N}{\mathbb{N}}

\newcommand{\R}{\mathbb{R}}
\newcommand{\C}{\mathbb{C}}
\renewcommand{\d}{\, \text{d}}

\newcommand{\trans}{\ensuremath{\mathsf{T}}}

\usepackage{color}

\usepackage{tikz}
\usepackage{pgfplots}

\usepackage[binary-units = true,exponent-product = \cdot]{siunitx}
\DeclareSIUnit{\vacuumpermeability}{\text{$\mu_0$}}

\journal{Journal of Magnetism and Magnetic Materials}

\begin{document}

\begin{frontmatter}



\title{Modeling the Magnetization Dynamics for Large Ensembles of Immobilized Magnetic Nanoparticles in Multi-dimensional Magnetic Particle Imaging}


\author[inst1]{Hannes Albers}
\author[inst2,inst3]{Tobias Knopp}
\author[inst2,inst3]{Martin M{\"o}ddel}
\author[inst2,inst3]{Marija Boberg}
\author[inst1,inst4]{Tobias Kluth}

\affiliation[inst1]{organization={Center for Industrial Mathematics, University of Bremen},
            addressline={Bibliothekstrasse 5}, 
            city={Bremen},
            postcode={28359},
            country={Germany}}

\affiliation[inst2]{organization={Section for Biomedical Imaging, University Medical Center Hamburg-Eppendorf},
addressline={Lottestrasse 55},
postcode={22529},
city={Hamburg},
country={Germany}}

\affiliation[inst3]{organization={Institute for Biomedical Imaging, Hamburg University of Technology},
addressline={Am Schwarzenberg-Campus 1},
postcode={21073},
city={Hamburg},
country={Germany}}

\affiliation[inst4]{organization={Center for Optimization and Approximation, University of Hamburg},
            addressline={Bundesstrasse 55}, 
            city={Hamburg},
            postcode={20146}, 
            country={Germany}}

\begin{abstract}
Magnetic nanoparticles (MNPs) play an important role in biomedical applications including imaging modalities such as magnetic resonance imaging (MRI) and magnetic particle imaging (MPI). The latter one exploits the non-linear magnetization response of a large ensemble of magnetic nanoparticles to magnetic fields which allows determining the spatial distribution of the MNP concentration from measured voltage signals. The image-to-voltage mapping is linear and described by a system matrix. Currently, modeling the voltage signals of large ensembles of MNPs in an MPI environment is not yet accurately possible, especially for liquid tracers in multi-dimensional magnetic excitation fields. As an immediate consequence, the system matrix is still obtained in a time consuming calibration procedure. While the ferrofluidic case can be seen as the typical setting, more recently immobilized and potentially oriented MNPs have received considerable attention. By aligning the particles magnetic easy axis during immobilization one can encode the angle of the particle's magnetic easy axis into the magnetization response providing a sophisticated benchmark system for model-based approaches. In this work we address the modeling problem for immobilized and oriented MNPs in the context of MPI. We investigate a model-based approach where the magnetization response is simulated by a N\'{e}el rotation model for the particle's magnetic moments and the ensemble magnetization is obtained by solving a Fokker-Planck equation approach. Since the parameters of the model are a-priori unknown, we investigate different methods for performing a parameter identification and discuss two different models: One where a single function vector is used from the space spanned by the model parameters and another where a superposition of function vectors is considered. We show that our model can much more accurately reproduce the orientation dependent signal response when compared to the equilibrium model, which marks the current state-of-the-art for model-based system matrix simulations in MPI.

\end{abstract}


\end{frontmatter}


\section{Introduction}

    
    
    Magnetic nanoparticles (MNP) play an important role in various biomedical applications. They can be used for diagnosis in imaging applications and therapy~\cite{krishnan2010biomedical}. Coupled with a drug they can be used for targeted drug delivery \cite{le2017real,zhang2017development,griese2020simultaneous}, e.g. to treat a vessel occlusion. In hyperthermia applications the particles are heated locally by applying an oscillating magnetic field, which can be used in tumor therapy. The treatment using MNP usually requires a good localization of the particles, which requires tomographic imaging techniques. One indirect way is to use magnetic resonance imaging (MRI) where the MNP shorten the $T_2^*$ relaxation time and in turn can be identified by a negative contrast mechanism. However, in practice it can be challenging to differentiate tissue and particle signal using MRI. In 2005 an alternative method called magnetic particle imaging (MPI) was proposed that exploits the non-linear magnetization response of magnetic nanoparticles to external magnetic fields~\cite{knopp2017magnetic}. Since these particles saturate even for low field strength in the region of only few mT$\mu_0^{-1}$, MPI does provide positive contrast without any background signal from tissue.
    
    MPI has shown to be a promising tool for a variety of medical application including vascular~\cite{herz2017magnetic,kaul2018magnetic,vogel2020superspeed} and oncological applications~\cite{zhu2019quantitative}. In particular MPI shows great potential in the area of intervention imaging since one can use one tracer to visualize the vessel tree using a blood pool tracer~\cite{khandhar2017evaluation,kaul2017vitro} and at the same time track medical devices such as guide wires, catheters~\cite{salamon2016magnetic,rahmer2017interactive}, and stents~\cite{herz2019magnetic}. In order to do so the particles are usually immobilized such that they can be permanently mounted on the device, e.g., by using a lacquer or by directly integrating the MNP into the polymer of a catheter. Quite recently, it has been shown that MPI can not only be used for tracking the position of MNP but also to determine the orientation~\cite{Moeddel2020IWMPI}. This is done by immobilizing the particles within a strong static magnetic field, which due to the MNP's magnetic anisotropy results in a particle ensemble with parallel aligned magnetic easy axis, which influences the magnetization response during excitation. Using multi-contrast image reconstruction techniques~\cite{rahmer2015first} it is then possible to simultaneously determine the position and orientation of a sample with immobilized easy axis aligned particles~\cite{MoeddelGrieseKluthKnopp2021_preprint}.
    
    \begin{figure*}[t]
        \centering
        \includegraphics[width=0.99\textwidth]{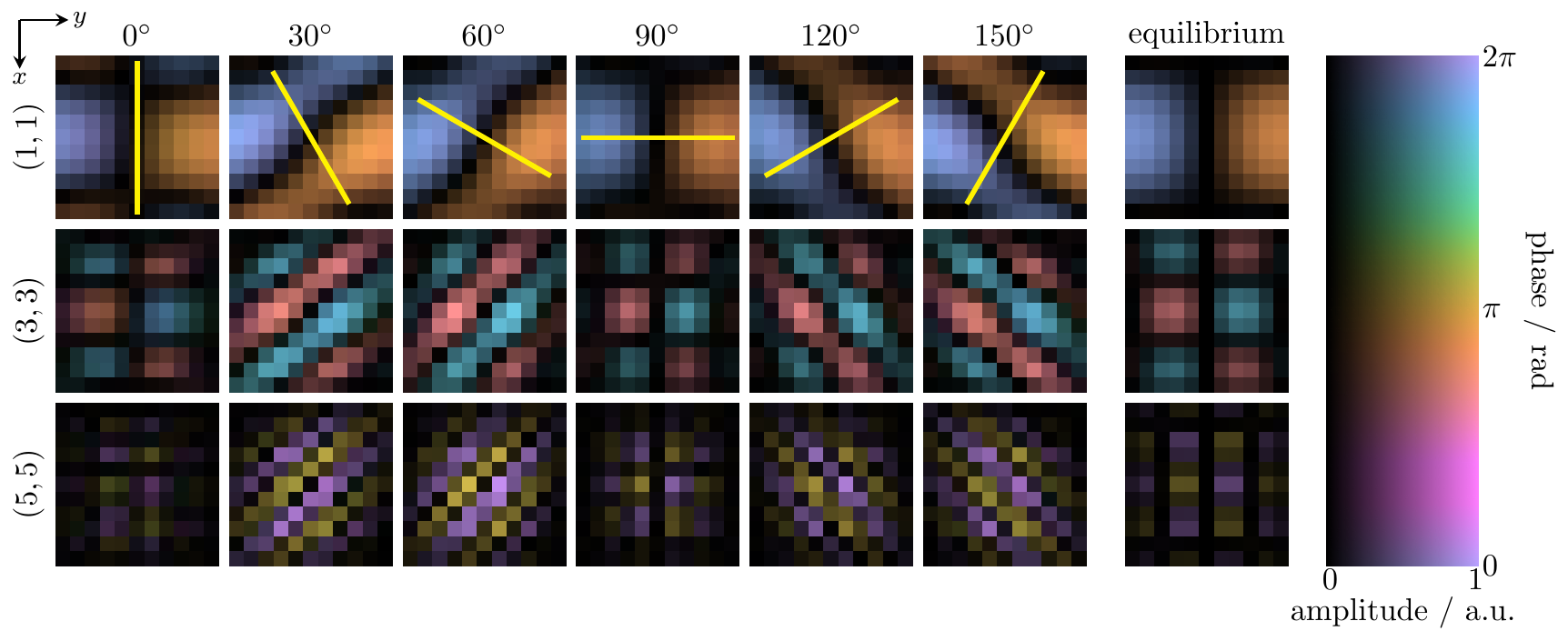}
        \caption{Here, the influence of the spatial orientation (parallel to the yellow line) of the easy axis of immobilized magnetic nanoparticles with parallel aligned magnetic easy axes on a 2D Lissajous type MPI system matrix is illustrated. Three selected frequency components with mixing factors $(m_x, m_y) \in \{ (1,1), (3,3), (5,5)\}$ for the $x$-channel are shown. On the left, system matrices measured for $\vartheta = \SI{0}{\degree},\SI{30}{\degree}, \dots, \SI{150}{\degree}$ are shown. On the right a fit of the equilibrium model (see M0 in Section~\ref{sec:data_driven_models}) is shown for particle diameter $D=\SI{22}{nm}$. The later is a good approximation for the system matrix components with $0^\circ$ and $90^\circ$ easy axis orientation only. In all figures the complex values of the system matrix components are colored using a 2D colormap shown on the right, where saturation encodes the amplitude (absolute value) and color encodes the phase of a complex number. Amplitudes are normalized for each frequency component as indicated by the a.u. to account for their wide range. This normalization remains the same along the different orientations and the equilibrium model.
        }
        \label{fig1:intro_example}
    \end{figure*}
    
    The image reconstruction problem in MPI consists of solving an inverse problem that is linear in the particle concentration~\cite{Knopp2010PhysMedBio}. The linear system obtained after discretization of the continuous imaging equation is ill-conditioned and requires proper regularization for a stable solution. Both the single-contrast and the multi-contrast imaging scenario can be formulated as a linear system and solved the same way using direct or iterative solvers. A prerequisite for image reconstruction is that the MPI system matrix $S\in \mathbb{C}^{K\times N}$ is known. The state-of-the art method for determining the system matrix $S$ is a fully data-driven calibration approach, where $S$ is measured column by column by measuring a small delta sample at each image voxel position. This procedure yields a physically accurate but noisy $S$ but it requires calibration times up to several days~\cite{grosser2020using} and it is also very memory demanding, particularly when considering 3D multi-patch field of views. 
    
    For this reason there is a series of research works aiming to replace this measurement-driven approach by a model-driven approach. In~\cite{knopp2009model} and~\cite{knopp20102d} it has been shown that the simple equilibrium particle magnetization model (see, e.g.,~\cite{Kluth2018a} for a definition) can be used to approximate the behavior of fluid MNP. The fully data-driven calibration method however, still outperformed these approaches for multi-dimensional MPI imaging protocols in terms of reconstruction image quality. In part this is due to the fact that the simplified equilibrium model relies on the assumption that the magnetization instantaneously follows the external magnetic field, i.e. no relaxation is present. In general, these assumptions are not fulfilled, especially when sufficiently fast changing fields are considered. In~\cite{KluthSzwargulskiKnopp2019} it has been shown that a more sophisticated magnetization model yields much more accurate system matrices for fluid MNP that allows to describe relaxation effects that can be observed in real experiments.
    
    In case of fast changing magnetic fields the dynamic behavior of the MNPs' magnetic moments is affected by Brownian (magnetic moment rotation due to rotation of the whole particle) and N\'{e}el (internal rotation of magnetic moment) mechanisms (see also~\cite{Coffey1992,shliomis1994theory}). For MPI, these new modeling approaches were initially suggested in~\cite{Weizenecker2010particle} in terms of stochastic ordinary differential equations and the first solutions using the Fokker-Planck equation for Brownian rotation in a one-dimensional setup were presented by Yoshida and Enpuku~\cite{Yoshida2012} followed by several works in this direction, e.g., see~\cite{weizenecker2012micro,Yoshida2012b,yoshida2013characterization,rogge2013simulation,reeves2014approaches,Kluth2018a} for further reading and~\cite{Martens2013,Deissler2014,Enpuku2014,Shah2015,graeser2015trajectory} for a stronger focus on MPI excitation patterns in the context of Brownian and N\'{e}el rotations. The N\'{e}el rotation was exploited to formulate the already mentioned sophisticated magnetization models for ferrofluids in~\cite{KluthSzwargulskiKnopp2019} where the authors also suggested an oriented easy axis model case which turned out to be not suitable for the fluid case. In particular, the N\'{e}el rotation mechanism is further influenced by the orientation of the particle's easy axis, which significantly affects the magnetization behavior, e.g., in the immobilized case for one-dimensional excitation, see~\cite{Yoshida2017,elrefai2021effect}. Even if the immobilized case is less complex when compared to the ferrofluid case where a coupling between Brownian and N\'{e}el rotation needs to be taken into account, the modeling problem of the magnetization of large ensembles of MNPs has not been addressed for multi-dimensional excitation patterns so far and is the major focus of the present work.

    \subsection{Problem Statement}
    
        In this work we consider immobilized magnetic nanoparticles with parallel aligned magnetic easy axes and aim at modeling MPI system matrices for different spatial orientations of said axis using the N\'{e}el rotation model B2 derived in~\cite{KluthSzwargulskiKnopp2019}. For orientation imaging~\cite{Moeddel2020IWMPI}, one needs to determine several system matrices $S^{\vartheta_l}$, $l=1,\dots, L$, where each matrix has been acquired under a different orientation angle $\vartheta_l$. Within this work we focus on easy axis alignments which lie within a plane. While for single-contrast imaging the data-driven calibration approach for determining the system matrix can still be handled, in the case of multi-contrast imaging it becomes too tedious since the calibration time is increased by a factor of $L$. For example, if the measurement of a single system matrix requires one day, as reported in~\cite{grosser2020using}, the multi-contrast calibration would require $L$ days in which the MPI system is occupied. A problem that is simply not an issue if a model-based approach is used to obtain the system matrices.
        
        Fig.~\ref{fig1:intro_example} shows a subset of measured MPI system matrices measured with an immobilized and easy-axis aligned sample at different orientations. We note that the image resolution is chosen small ($11 \times 11$) such that the calibration time could be kept short. The measured matrices serve as the ground truth in the present work and the details of the experimental parameters are outlined in Section~\ref{sec:experimental_setting}. The image shows three selected matrix rows that have been reshaped to represent the signal that a certain spatial position contributed to the voltage signal in Fourier space within a tomographic imaging experiment. This representation is very well known in MPI since the matrix rows for the equilibrium model can be represented using tensor products of Chebyshev polynomials~\cite{maass2020representation} for Lissajous-type excitations. What is much more interesting now and has not been studied is the influence of parallel aligned magnetic easy axes on the structure of the MPI system matrices. Independent of the orientation of this axis one can see that a wave-hill structure is preserved in this case. However, what is also visible is that the wave hills are merging along certain directions which are global for each easy axis orientation. When changing the orientation, the merging angle changes accordingly, which provides the contrast mechanism for orientation imaging. Another observation is an increased amplitude when the orientation is closer to $45^\circ$ or $135^\circ$. Here, it is conspicuous that these angles are the orientations of the drive field vector in which a fast change of the length of the applied magnetic field vector is performed in the used cosine Lissajous-type applied magnetic field. 
        
        What one can also see is that the equilibrium model is not expressive enough to describe the observed effect at all and thus, this model is not suitable for modeling the system matrices in this particular multi-contrast experiment. The purpose of the present work is to use the model developed in~\cite{KluthSzwargulskiKnopp2019} and calculate model-based system matrices that can describe the easy-axis dependent effects on the MPI system matrix shown in Figure~\ref{fig1:intro_example}. This is not only an important prerequisite for the development of more efficient calibration procedures but also for future investigations of different applied fields and their optimization with respect to multi-contrast imaging for magnetic easy axis orientation.

\section{Materials and Methods}

Motivated by a general MPI imaging experiment where the particles are located in the MPI scanner and the change of the particle magnetization is measured using induction coils, we outline the general setting in the following. Let $\Omega \subset \R^3$ be the field-of-view (below we further restrict $\Omega$ to be the $xy$-plane). A voltage induced in a receive coil with sensitivity profile $p: \R^3 \rightarrow \R^3$ in m$^{-1}$ is given by
\begin{align}
 \tilde{v} (t) &=- \mu_0 \int_\Omega  c(r) p(r)^\trans \frac{\partial}{\partial t}\bar{m}(r,t) \ \d r   \label{eq:complete-problem}
\end{align}
in V where $c: \Omega \rightarrow \R^+_0$ in \si{\mol\per\liter} is the concentration of the magnetic nanoparticles and $\bar{m}: \R^3 \times [0,T] \rightarrow \R^3$ in $10^{-3}$\si{\ampere\meter\squared\per\mol} is the molar mean magnetic moment. 
The mean magnetic moment $\bar{m}(r,t)$ depends on the applied magnetic field $H: \R^3 \times [0,T] \rightarrow \R^3$ in \si{\tesla\per\vacuumpermeability}, which is usually a $T$-periodic function represented on the time interval $[0,T]$. In MPI a static selection field $H_\mathrm{SF}: \R^3 \rightarrow \R^3$ is combined with a dynamic drive field $H_\mathrm{DF}: \R^3 \times [0,T]\rightarrow \R^3$, i.e., $H(r,t)=H_\mathrm{SF}(r) + H_\mathrm{DF}(r,t)$. Note that the notation is motivated by the MPI point of view. Effectively, the selection field evaluated at discrete points encodes a variety of offset fields taken into account. The induced signal is analogously filtered to remove the main contribution of the direct feed-through. In good approximation it is represented by a convolution with a $T$-periodic filter kernel $a:\R \rightarrow \R$ yielding the signal $v=a\ast\tilde{v}$. Therefore, $v$ is $T$-periodic as well and can be expanded into a Fourier series with coefficients
\begin{align}
 \hat{v}_k &= - \hat{a}_k \frac{\mu_0}{T} \int_\Omega  c(r) p(r)^\trans \int_{0}^{T}  \left( \frac{\partial}{\partial t}\bar{m}(r,t) \right){\textrm{e}}^{-2\pi i tk/T} \ \d t \ \d r \label{eq:MPI_model_fourier}
\end{align}
for $k\in\N_0$ and the conjugate complex $\hat{v}_{-k}=\overline{\hat{v}_k}$.
This formulation is commonly used, since the signal at the excitation frequencies is blocked using an analog band-stop filter prior to the signal digitization. With $s_k(r) := - \hat{a}_k\frac{\mu_0}{T} \int_{0}^{T} p(r)^\trans \frac{\partial}{\partial t}\bar{m}(r,t) {\textrm{e}}^{-2\pi i tk/T} \ \d t$
we can bring this into the standard notation
\begin{align}
 \hat{v}_k &=\int_\Omega  c(r) s_k(r) \ \d r,  \label{eq:system_function_model}
\end{align}
where $s_k: \Omega \subset \R^3 \rightarrow \C$ is the system function, e.g., as previously discussed in Figure~\ref{fig1:intro_example}. 
When measuring the system function at discrete positions $r_\nu \in \Omega, \nu=1,\dots,N$, and with a certain sampling rate in time, one obtains the MPI system matrix $S = \left( s_k(r_\nu) \right)_{k=0,\dots,K; \nu=1,\dots,N}$ storing the Fourier transformed measurements in the respective columns. In this work we  consider both modeled system matrices as well as measured system matrices that are obtained by a hybrid approach~\cite{von2017hybrid} exploiting the focus field in the MPI scanner. 
In case of multiple receive channels, the corresponding system matrices can be stacked to arrive at a joint linear system of equation; see for example~\cite{Kluth2019numerical} for a formal definition.

\subsection{Particle magnetization model} 
%

One important aspect in the model equation \eqref{eq:MPI_model_fourier} is the mean magnetic moment $\bar{m}$ of the ensemble of nanoparticles. 
Over-simplified models like the \textit{equilibrium model} (see, e.g.,~\cite{Kluth2018a}) do not take into account the particles' anisotropy.
As we have the immobilized and oriented particle configuration in mind, we need a more sophisticated model and  exploit a N\'{e}el rotation model taking into account an uniaxial anisotropy. 
For this we follow the Fokker-Planck equation approach for the Landau-Lifshitz-Gilbert equation as already described in~\cite{KluthSzwargulskiKnopp2019}.
We determine the mean magnetic moment via the probability density function $f: \Omega \times \mathbb{S}^2 \times [0,T] \rightarrow \R_0^+$ which is the solution to the corresponding Fokker-Planck equation where $\mathbb{S}^2$ is the surface of the sphere in $\R^3$.
The mean is then given by
\begin{equation}\label{eq:mean_magmom_FP}
  \bar{m}(r,t) = m_0 \int_{\mathbb{S}^2} m f(r,m,t) \ \d m
\end{equation}
where $f$ is the solution to the following specific case of a convection-diffusion equation on the sphere
\begin{equation}
\label{eq:FP-general}
 \frac{\partial}{\partial t} f = \mathrm{div}_{\mathbb{S}^2}\left(\frac{1}{2\tau} \nabla_{\mathbb{S}^2} f \right) - \mathrm{div}_{\mathbb{S}^2}\left(b f\right)
\end{equation}
where $\tau >0$ is the relaxation time constant and the (velocity) field $b:\mathbb{S}^2 \times \R^3 \times \mathbb{S}^2 \rightarrow \R^3$ given by
\begin{align}
 &b(m,H,n) = p_1 H \times m + p_2 (m\times H) \times m \notag \\ & \quad+ p_3 (n\cdot m) n \times m + p_4 (n\cdot m) (m\times n) \times m \label{eq:convection}
\end{align}
where $p_i\geq 0$, $i=1,\hdots,4$, are physical constants and $n\in \mathbb{S}^2$ is the easy axis of the particles. 
Differentiation in terms of gradient $\nabla_{\mathbb{S}^2} $ and divergence $ \mathrm{div}_{\mathbb{S}^2}$ is considered with respect to the surface $\mathbb{S}^2$.
The Euclidean scalar product of $\R^3$ is denoted by $n\cdot m$.
  A more detailed derivation of the Fokker-Planck equation from the Langevin equation can, for example, be found in~\cite{Kluth2018a}. 
 A N\'{e}el rotation including anisotropy is then given by $p_1=\tilde{\gamma}\mu_0$, $p_2=\tilde{\gamma}\alpha \mu_0$, $p_3=2\tilde{\gamma}\frac{K^\text{anis}}{M_\mathrm{S}}$, and $p_4=\alpha p_3$ with $\tau=\frac{V_\mathrm{C} M_\mathrm{S}}{2 k_\mathrm{B} T_\mathrm{B} \tilde{\gamma} \alpha}$ and $\tilde{\gamma}=\frac{\gamma}{1+ \alpha^2}$. Here, $V_\mathrm{C}$ is the core volume of the nanoparticles depending on the core diameter $D$, $M_\mathrm{S}$ denotes the saturation magnetization, $\alpha$ is the damping parameter and $k_\mathrm{B}$ and $T_\mathrm{B}$ denote the Boltzman constant and the temperature, respectively. The uniaxial anisotropy constant is denoted by $K^\text{anis}$.
 We note that the parabolic partial differential equation in \eqref{eq:FP-general} has no dependence on derivatives with respect to the spatial variable $r$.
 It can thus be considered as parametric with respect to $r$, respectively the constant offset fields encoded in the selection field.

The equation is solved numerically by using a finite volume method, which is conservative by design in the sense that the integral over the solution does not change over time. This makes sense, since the solution of the Fokker-Planck equation is a probability density. The sphere is segmented into spherical triangles, and considering the probability flow over the edges leads to a system of ordinary differential equations. This system is subsequently solved with a multistep integrator. Here, we exploited the computational toolbox provided in~\cite{AlbersKluthKnopp2020_preprint}, where more details on the implementation can be found.

\subsection{Data-driven physically motivated models}\label{sec:data_driven_models}

Given an appropriate model for the mean magnetic moment $\bar{m}$ it is still unlikely that single scalar parameters such as particle diameter and anisotropy constant, which would represent a perfect monodisperse tracer, will be sufficient for the proper prediction of the measured signal in the MPI setting. 
 Additionally, in an MPI measurement we do not have direct access to the mean magnetic moment such that we need to take into account uncertainties in the involved quantities describing the system function $s_k$ in \eqref{eq:system_function_model}. 
We thus formulate the following physical model-based but still data-driven approaches for the system functions. 

The general setting is based on the following assumptions:
\begin{itemize}
\item[(i)] The analog filter $\hat{a}$ is not given with necessary accuracy.
\item[(ii)] The sensitivity profile $p$ of the receive coil is homogeneous and it is assumed to be given.  
\item[(iii)] The monodisperse N\'{e}el rotation model $\bar{m}_{(D,K^\text{anis},\vartheta)}$ including uniaxial anisotropy constant $K^\text{anis}$, particle diameter $D$, and easy axis $n(\vartheta)=(\cos (\vartheta),\sin(\vartheta),0)^\trans\in S^2$ according to \eqref{eq:mean_magmom_FP}, \eqref{eq:FP-general}, \eqref{eq:convection}, i.e., $\bar{m}_{(D,K^\text{anis},\vartheta)}=\bar{m}$ with $\bar{m}$ from \eqref{eq:mean_magmom_FP}, is exploited in the data-based approaches in the following. 
We particularly consider easy axes in the $xy$-plane for a given angle $\vartheta$ with respect to the $x$-axis.
\end{itemize}
With these assumptions, predefined discrete parameter sets $\mathcal{P}_\text{diam},\mathcal{P}_\text{anis}, \mathcal{P}_\text{angle}$, and a frequency index set $I_\text{freq}\subset\{0,1,\hdots,K\}$  we consider the following data-driven models describing the system functions in \eqref{eq:system_function_model}
\begin{itemize}
\item[M0] Even if the equilibrium model~\cite{Kluth2018a} is not able to express the dependence on the easy axis orientation $\vartheta$, we include it as a reference to the state of the art in MPI and also for the sake of completeness: For individual $D\in \mathcal{P}_\text{diam}$ and $\vartheta\in \mathcal{P}_\text{angle}$ identify $\hat{a}=(\hat{a}_k)_{k\in I_\text{freq}}$ in
    \begin{equation}
    s^{(\vartheta)}_k(r)= \hat{a}_k \hat{\phi}_{D;k}(r), \ k\in I_\text{freq}    
    \end{equation}  
    for given  $\hat{\phi}_{D;k}(r)=\frac{1}{m_0 T}\int_{0}^{T} p^\trans \frac{\partial}{\partial t}\bar{m}^\text{equ}_{D}(r,t) {\textrm{e}}^{-2\pi i tk/T} \ \d t$ from measurements $S^{\mathrm{calib}}_{\vartheta}$ (measured system matrix for one predefined easy axis angle $\vartheta$; see Section~\ref{sec:experimental_setting}). 
    $\bar{m}^\text{equ}_{D}$ is explicitly given by the equilibrium model , i.e, 
\begin{equation}
\bar{m}^\text{equ}_{D}(r,t) = m_0 \mathcal{L}_\beta(| H(r,t)|) \frac{H(r,t)}{| H(r,t)|} 
\end{equation}
where $\mathcal{L}_\beta: \R \rightarrow \R$ is given in terms of the Langevin function by the following:
\begin{equation}
\label{eq:Langevin}
 \mathcal{L}_\beta(z) =  \left( \coth( \beta z) - \frac{1}{ \beta z} \right)
\end{equation}
for $m_0,\beta >0$. Both $m_0=V_\mathrm{C} M_\mathrm{S}$ and $\beta=\frac{\mu_0 V_\mathrm{C} M_\mathrm{S}}{k_\mathrm{B} T_\mathrm{B}}$ with $V_\text{C} = \frac{1}{6}\pi D^3$ depend on the particle core diameter $D$ in a cubic fashion.
    \item[M1] For individual $D\in \mathcal{P}_\text{diam}$, $K^\text{anis}\in \mathcal{P}_\text{anis}$, and $\vartheta \in \mathcal{P}_\text{angle}$ identify $\hat{a}=(\hat{a}_k)_{k\in I_\text{freq}}$ in
    \begin{equation}
    s^{(\vartheta)}_k(r)= \hat{a}_k \hat{\phi}_{(D,K^\text{anis},\vartheta);k}(r), \ k\in I_\text{freq}    
    \end{equation}  
    for given  
    \begin{equation}
     \hat{\phi}_{(D,K^\text{anis},\vartheta);k}(r)=\frac{1}{m_0 T}\int_{0}^{T} p^\trans \frac{\partial}{\partial t}\bar{m}_{(D,K^\text{anis},\vartheta)}(r,t) {\textrm{e}}^{-2\pi i tk/T} \ \d t   
    \end{equation}
    from measurements $S^{\mathrm{calib}}_{\vartheta}$ (measured system matrix for one predefined easy axis angle $\vartheta$; see Section~\ref{sec:experimental_setting}).
    \item[M2] In the absence of an accurate estimate of the transfer function and aiming for a polydisperse model, as a preparatory step we generate an improved dictionary of functions such that we follow the following heuristic two step procedure: 
    \begin{enumerate}
        \item For any $D\in \mathcal{P}_\text{diam}$, $K^\text{anis}\in \mathcal{P}_\text{anis}$, and $\vartheta \in \mathcal{P}_\text{angle}$ obtain an $\hat{a}_{(D,K^\text{anis},\vartheta)}$ according to the procedure outlined for M1 which gives us the preprocessed auxiliary functions 
        \begin{equation}\label{eq:psi_def}
               \psi_{(D,K^\text{anis},\vartheta);k}(r)= \hat{a}_{(D,K^\text{anis},\vartheta);k} \hat{\phi}_{(D,K^\text{anis},\vartheta);k}(r), \ k\in I_\text{freq}.     
        \end{equation}
        \item  We then obtain a weighted linear combination with respect to the physical parameters to take the polydisperse characteristic of the tracer into account. 
                        Here, we aim for finding weights $w_{D,K^\text{anis}}\geq 0$ such that
            \begin{equation}
                s^{(\vartheta)}_k(r)=\negthickspace\!\sum_{D\in\mathcal{P}_\text{diam}} \sum_{K^\text{anis}\in\mathcal{P}_\text{anis}}  w_{D,K^\text{anis}} \psi_{(D,K^\text{anis},\vartheta );k}(r), \ k\in I_\text{freq} 
            \end{equation}
               approximates the measurements $S^{\mathrm{calib}}_{\vartheta}$ for any $\vartheta \in \mathcal{P}_\text{angle}$.
               In contrast to M1 we obtain global weights not depending on $\vartheta$ but only depending on $D$ and $K^\text{anis}$.
    \end{enumerate}
\end{itemize}

\textit{Remark:} Note that the fitted $\hat{a}$/$\hat{a}_{(D,K^\text{anis},\vartheta)}$ are parallel to the transfer function in \eqref{eq:MPI_model_fourier} but its absolute also includes some scalar factors (such as $\mu_0$, $m_0$, etc.).

\paragraph{Technical details regarding the parameter identification}

We exploit the discrete nature in space given by the discrete points $\{r_\nu\}_{\nu=1,\hdots,N} \subset \Omega$ of the problem setup and introduce the following variables and transformations:
\begin{itemize}
    \item $S^{\mathrm{model}} = (s_k^{(\vartheta)}(r_\nu))_{k\in I_\text{freq},\nu=1,\hdots,N, \vartheta \in \mathcal{P}_\text{angle}} \in \C^{|I_\text{freq}| \times N \times |\mathcal{P}_\text{angle}|}$ being a tensor including system matrices for the respective model for different orientations $\vartheta$. With $S^\mathrm{model}_\vartheta$/$S^\mathrm{calib}_\vartheta \in \C^{|I_\text{freq}| \times N}$ we denote the respective system matrix for easy axis orientation $\vartheta$.
    \item $\Phi_{(D,K^\text{anis},\vartheta)}=(\hat{\phi}_{(D,K^\text{anis},\vartheta);k}(r_\nu))_{k\in I_\text{freq},\nu=1,\hdots,N} \in \C^{|I_\text{freq}| \times N}$ being an element of the dictionary of system matrices according to the description of M1. $\Phi_D$ analogously for M0.
    \item $\Psi_{(D,K^\text{anis},\vartheta)}=({\psi}_{(D,K^\text{anis},\vartheta);k}(r_\nu))_{k\in I_\text{freq},\nu=1,\hdots,N} \in \C^{|I_\text{freq}| \times N}$ being an element of the dictionary of system matrices according to \eqref{eq:psi_def}
    \item $T^\mathrm{mat2vec}: \C^{K\times N} \rightarrow \C^{NK}$ concatenates the columns of the input matrix in one vector.
    \item 
    $T^\mathrm{diag}: \C^K \rightarrow \C^{K\times K}$ generates a diagonal matrix with the input vector as the diagonal.
\end{itemize}

Model fitting in any of the previously outlined model approaches is performed by doing a (possibly weighted) least squares fit to the data.

\begin{itemize}
    \item[M0/M1] For any $\vartheta \in \mathcal{P}_\text{angle}$ and 
    \begin{equation*} 
        q:=\left\{ \begin{array}{ll} D \in \mathcal{P}_\text{diam} & \text{for M0} \\ (D,K^\text{anis},\vartheta) \in \mathcal{P}_\text{diam} \times \mathcal{P}_\text{anis} \times \{\vartheta\} & \text{for M1}  \end{array} \right.
    \end{equation*}
    we obtain 
    \begin{align}
    \hat{a}_q &= \mathrm{arg\ min}_{\hat{a}\in \C^{|I_\text{freq}|}} \| T^\mathrm{diag}(\hat{a}) \Phi_q - S^{\mathrm{calib}}_{\vartheta} \|_\mathrm{F}^2 \notag \\ &\left(= \mathrm{arg\ min}_{\hat{a}\in \C^{|I_\text{freq}|}} \| T^\mathrm{mat2vec}(T^\mathrm{diag}(\hat{a}) \Phi_q - S^{\mathrm{calib}}_{\vartheta}) \|_2^2\right)   
    \end{align}
    explicitly by (see, e.g.,\cite[Eq. (21)]{Kluth2017})
    \begin{equation}
    \hat{a}_{q;k} = \frac{S^{\mathrm{calib}}_{\vartheta;k} \overline{\Phi_{q;k}}^T}{\| \Phi_{q;k} \|_2^2}
    \end{equation}
    for any $k \in \{1, \hdots,|I_\text{freq}|\}$ where $S^{\mathrm{calib}}_{\vartheta;k}$ and $\Phi_{q;k}$ denote the respective $k$-th row.  
\item[M2] 
The first step is given in terms of  $\Psi_q=T^\mathrm{diag}(\hat{a}_q) \Phi_q,\ q \in  \mathcal{P}_\text{diam} \times \mathcal{P}_\text{anis} \times \mathcal{P}_\text{angle}$ where $\hat{a}_q$ are obtained as outlined in M1.
    In the second step, we utilize for any $\vartheta \in \mathcal{P}_\text{angle}$ and $q^{(\vartheta)}_j:=(D_j,K_{\mathrm{anis};j},\vartheta) \in \mathcal{P}_\text{diam} \times \mathcal{P}_\text{anis} \times \{ \vartheta \}$, $j=1,\hdots,J:=|\mathcal{P}_\text{diam}|\cdot |\mathcal{P}_\text{anis}|$, (such that $\{q_j^{(\vartheta)}\}_{j=1,\hdots,J}=\mathcal{P}_\text{diam} \times \mathcal{P}_\text{anis} \times \{\vartheta\}$), we obtain $w\in \R_+^J$ by solving
    \begin{equation}\label{eq:LSfitM2}
        \min_{w\in \R^J_+} \sum_{\vartheta \in \mathcal{P}_\text{angle}} \| M_{\vartheta}w - T^\mathrm{mat2vec}(\Sigma_{\vartheta} S^{\mathrm{calib}}_{\vartheta}) \|_2^2 
    \end{equation}
    with dictionary matrices $$M_{\vartheta}:=\left[T^\mathrm{mat2vec}(\Sigma_{\vartheta}\Psi_{q_1^{(\vartheta)}}), \hdots, T^\mathrm{mat2vec}(\Sigma_{\vartheta} \Psi_{q_J^{(\vartheta)}}) \right]$$
    for $\vartheta \in \mathcal{P}_\text{angle}$ which includes the simulations at different positions in each column. Each column corresponds to one physical parameter tuple $q_j^{(\vartheta)}$. $\Sigma_{\vartheta} \in \R^{|I_\text{freq}| \times |I_\text{freq}|}$ is a diagonal weighting matrix with respect to the individual frequency components allowing a weighted least squares fit.
    
    Throughout this work, the weighting was chosen such that for each frequency, the measurement is normalized with respect to the infinity-norm over the offset fields:
        $$\Sigma_{\vartheta} = T^\mathrm{diag}\left(\left( \| S^{\mathrm{calib}}_{\vartheta;k}\|_\infty^{-1}\right)_{k\in I_\text{freq}}\right).$$
\end{itemize}

For numerical treatment, the least-squares functional is expanded, taking into account that the data is complex while the weights are real. Thus, we obtain a so-called quadratic program with linear constraints of the form $\min_w \frac{1}{2}w^T A w + f^Tw$ for a positive semidefinite real matrix $A$ and a real vector $f$, with the constraints $w_i \geq 0$ for $i=1,\dots,J$. We solve this problem by using a standard interior-point algorithm, as described in~\cite{optimization}, among many other works.

However, since a rather high resolution of parameters was considered for the data fitting in model M2, many matrix columns of the resulting matrices $M_{\vartheta}$ differ only slightly. This leads to an ill-conditioned problem and as a result, the weight vector $w$ is not interpretable, since it might simply reflect noise in the measurements. This makes it necessary to employ regularization techniques in order to obtain weights that better reflect the real-world parameter distribution. We chose to employ the truncated singular value decomposition (TSVD) or reduced-rank regularization approach which is motivated by the equivalent formulation of \eqref{eq:LSfitM2} as 
    \begin{equation}\label{eq:LSfitM2_alternative}
        \min_{w\in \R^J_+}  \| Mw - \underset{=:y}{\underbrace{(T^\mathrm{mat2vec}(\Sigma_{\vartheta} S^{\mathrm{calib}}_{\vartheta}))_{\vartheta\in \mathcal{P}_\text{angle}}}} \|_2^2,
    \end{equation}
where the matrix $M\in \C^{|I_\text{freq}|\cdot |\mathcal{P}_\text{angle}| \times |\mathcal{P}_\text{diam}| \cdot |\mathcal{P}_\text{anis}|}$ consists of the stacked $M_{\vartheta}, \vartheta \in \mathcal{P}_\text{angle}$. For this we compute the SVD, i.e., $M = U \Sigma V^\ast$ for unitary matrices $U, V$ and a real diagonal matrix $\Sigma$. Choosing a low-rank approximation by the SVD is particularly motivated by the following relationship 
\begin{align}
    \| M w -y \|_2 &= \| U^\ast M w - U^\ast y  \|_2 \notag \\
    &\approx \| U_s^\ast M w - U_s^\ast y \|_2, \label{eq:approx_low_rank}
\end{align}
where $U_s := (U_{ij})_{i=1,\hdots,|\mathcal{P}_\text{angle}|\cdot|I_\text{freq}|; j=1,\hdots,s}$ includes the first $s$ columns of $U$, corresponding to the $s$ largest singular values, i.e., $s$ being the regularization parameter.


We thus obtain the regularized minimization problem
\begin{align*}
    \min_{w \in \mathbb{R}^J_+} \| U_s^\ast M w - U_s^\ast y\|_2^2.
\end{align*}
Note that the global error measure (see Section~\ref{sec:distance_measure}) $\varepsilon_\mathrm{global}(S^\mathrm{model},S^\mathrm{calib}) \sim \| M w -y \|_2$. Due to the fast decay of singular values, using a low-rank approximation thus has only little influence on the global error measure but stabilizes the parameter reconstruction.


\subsection{Distance Measure} \label{sec:distance_measure}
We exploit the distance measure as introduced in~\cite{KluthSzwargulskiKnopp2019}. 
For quantifying the accuracy of a model 
we express the frequency index $k$ in terms of the mixing orders $m_x$, $m_y$~\cite{rahmer2012analysis} fulfilling
\begin{align} \label{Eq:mix}
    k(m_x,m_y) &= m_x N^{\text{dens}} + m_y (N^{\text{dens}}+1)
\end{align}
Here, $N^{\text{dens}}$  is the parameter that controls the density of the Lissajous sampling pattern and for each $k$ the factors $m_x$ and $m_y$ with the smallest absolute value fulfilling \eqref{Eq:mix} are considered.
Now with $S^{\text{calib}}_{\vartheta;k(m_x,m_y)}$ being the measured and $S^{\text{model}}_{\vartheta;k(m_x,m_y)}$ being the modeled $k(m_x,m_y)$-th system matrix row (also named frequency component) for a given angle $\vartheta$ we consider the normalized root mean square error (NRMSE)
\begin{align} \label{Eq:error1}
    \varepsilon(S^{\text{calib}}_{\vartheta;k}, S^{\text{model}}_{\vartheta;k}) &= \frac{\frac{1}{\sqrt{N}}\Vert  S^{\text{calib}}_{\vartheta;k}- S^{\text{model}}_{\vartheta;k} \Vert_2}{\Vert  S^{\text{calib}}_{\vartheta;k} \Vert_\infty},
\end{align}
where we decided to use the infinity norm for normalization.
In addition to this per-row metric we consider global measures over $k \in I_\text{freq} =\{k(m_x,m_y)\,|\, m_x=0,\dots,M_x$, $m_y=0,\dots,M_y \}\setminus \{ (0,0),(0,1),(1,0)\}$ where $M_x,M_y\in\N$ are upper bounds for the mixing order and $\vartheta \in \mathcal{P}_\text{angle}$ with $\mathcal{P}_\text{angle}$ being a discrete set of orientations.
In particular we consider the 
\begin{itemize}
    \item per frequency NRMSE measure 
    \begin{align}
    &\varepsilon^\text{freq}_{k}(S^{\text{calib}}, S^{\text{model}}) \nonumber \\ &= \frac{1}{\sqrt{|\mathcal{P}_\text{angle}|}} \left( \sum_{\vartheta \in \mathcal{P}_\text{angle}} \varepsilon(S^{\text{calib}}_{\vartheta;k}, S^{\text{model}}_{\vartheta;k})^2 \right)^{\frac12}    
    \end{align}
    \item and the global NRMSE measure 
       \begin{align}
       &\varepsilon^\text{global}(S^{\text{calib}}, S^{\text{model}}) \notag \\ &= \frac{1}{\sqrt{|I_\text{freq}|}} \left( \sum_{k\in I_\text{freq}} \varepsilon^\text{freq}_{k}(S^{\text{calib}}, S^{\text{model}})^2 \right)^{\frac12}  \label{eq:globalMeasure}  
    \end{align}
 
\end{itemize}
  We note that in both measures the quadratic sums have been additionally performed along the receive channel dimension, which we omitted in this work to simplify the notation.

\subsection{Experimental Setup}
\label{sec:experimental_setting}


Experimental data is acquired using a pre-clinical MPI system (Bruker, Ettlingen, Germany) that is equipped with three send and three receive channels. The scanner is operated with a 2D excitation in the $xy$-plane (horizontal) with \SI{12.59}{\milli\tesla\per\vacuumpermeability} excitation field strength in $x$-direction and  \SI{13.55}{\milli\tesla\per\vacuumpermeability} in $y$-direction. These two values have been retrospectively determined using a calibration coil and thus do not exactly match the ones used at the scanner console. The drive-field frequencies are $f_x=2.5/102$ MHz and $f_y=2.5/96$ MHz leading to a repetition time of  \SI{652.8}{\micro\second} and a density of the Lissajous trajectory of $N^{\text{dens}} = 16$. 
For signal reception we use a 3D pick-up coil while taking only the two channels in $x$- and $y$-direction into account. The analog signal is band-pass filtered and digitized at a sampling rate of \SI{2.5}{\mega\hertz}.

The system matrices are sampled on a grid of size $11 \times 11$ at $N=121$ equidistant measurement points with different offset values ranging from 
\SI{-14.35}{\milli\tesla\per\vacuumpermeability} to \SI{14.35}{\milli\tesla\per\vacuumpermeability} in $x$-direction and \SI{-14.71}{\milli\tesla\per\vacuumpermeability} to \SI{14.71}{\milli\tesla\per\vacuumpermeability} in $y$-direction. Both intervals have been calibrated using a  3-channel gaussmeter with a three-axis Hall probe (model 460, Lake~Shore, Westerville, USA). Somewhat uncommonly we report the sampled grid using offset values since we measure the matrix with a static sample at the center and move the sampling trajectory using a 3D offset field that can be adjusted using the scanner. This method is called hybrid system matrix~\cite{von2017hybrid} and it is advantageous for the present study as spatial field imperfections can be ignored since the sample is always measured at the same position. Since the offset field in a real MPI imaging experiment would be translated to space, we consider the $x$- and $y$-offset field dimensions to be spatial dimensions with a virtual and perfect selection field applied. The overscan \cite{weber2015artifact} of the calibration data is 13.9~\% in the $x$-direction and 8.6~\% in the $y$-direction. All measurements have been done with an active gradient of \SIlist{-1.0;-1.0;2.0}{\tesla\per\vacuumpermeability\per\meter} in $x$\nobreakdash-, $y$-, and $z$-direction, which, however, is only enabled to suppress background signals that are much stronger with disabled selection field. This should not be confused with the virtually applied selection field gradient, which can have any value.
The calibration measurements are block averaged with a factor of \num{5000} amounting to a measurement time of about \SI{3.3}{\second} per position. 

\begin{figure}[t!]
    \centering
    \includegraphics{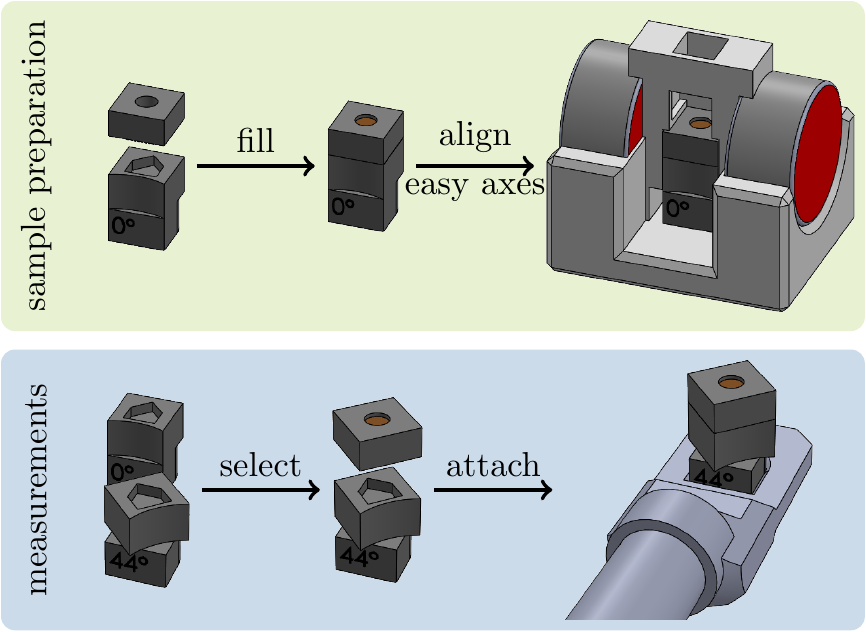}
    \caption{A schematic overview of delta sample (top) and measurement (bottom) preparation is shown. The delta sample is assembled from two parts (top left). Its cylindrical depression is filled with a brownish alginate-tracer mixture (top center). Immediately, the sample is placed in a magnetic field (red sided permanent magnets) to align the magnetic nanoparticles magnetic easy axis in parallel and immobilize particles by drying (top right). The sample has multiple connectors, which can be selected (bottom left). Each connector can be assembled with the top part of the sample (bottom center) and attached to the robot arm of our MPI system (bottom right), providing a connector specific easy axis orientation inside the MPI system.}
    \label{fig:phantom}
\end{figure}

For system matrix calibrations a single cylindrical delta sample of immobilized magnetic nanoparticles with magnetic easy axis in parallel to a specified direction is designed. As shown in Fig.~\ref{fig:phantom}, the delta sample is assembled from two parts, which are 3D printed using a Form 3 and clear resin (Formlabs Inc., Somerville, USA). The top part (carrier) has a \SI{3}{\milli\meter} diameter and \SI{3}{\milli\meter} height cavity to hold magnetic nanoparticles. The bottom part (connector) is interchangeable and can be assembled with the carrier. The assembled delta sample can be mounted onto the robot arm of our MPI system such that it has a defined orientation inside the $xy$-plane of the MPI system, which depends on the connector chosen. Initial preparation of the delta sample is done by assembling the carrier with a \SI{0}{\degree}-connector. The carrier is then filled with a mixture of \SI{20}{\micro\liter} fluidal perimag (Micromod Partikeltechnologie GmbH, Rostock, Germany) with an iron concentration of \SI{89}{\milli\mol\of{Fe}\per\liter} and about \SI{21}{\micro\liter} sodium alginate powder. At this point the nanoparticles can still change their spatial orientation by rotation. The sample is then immediately put in-between two cylindrical neodymium magnets with a \SI{0.4}{\tesla\per\vacuumpermeability} axial field, which was measured using the same gaussmeter as used for the MPI scanner offset calibration. The magnetic easy axes of the nanoparticles align in parallel to the external field~\cite{yoshida2017effect,Elrefai2020} and slowly lose their mobility due to drying of the mixture, which fixes the orientation of their easy axes permanently. Placed into our MPI system the delta sample with the \SI{0}{\degree}-connector has a $\vartheta = \SI{0}{\degree}$ angle between the parallel aligned magnetic easy axis and the $y$-axis of our MPI system.

In this preparation process the mixing of perimag and sodium alginate powder is most error prone, which is why only a single carrier is prepared and different orientations inside our MPI system are realized by disassembling carrier and connector and reassembling the former with a different suitable connector. For our measurements a series of connectors was printed, allowing us to measure \num{12} system matrices at angles 
\begin{equation}
 \mathcal{P}_\text{angle} := \left\lbrace \SI{0}{\degree}, \SI{15}{\degree},\dots,\SI{165}{\degree}\right\rbrace
\end{equation}
resulting in the measured system matrix tensor $S^\mathrm{calib} \in \C^{K\times N \times |\mathcal{P}_\text{angle}|}$.

\subsubsection{Simulation Setting}
\label{sec:simulation_setting}

For the parameter optimization we selected for each model parameter a finite set of values and performed simulations with them. For the particle diameter $D$ we chose 
\begin{equation}
    \mathcal{P}_\text{diam} = \left\lbrace \SI{14}{nm}, \SI{16}{nm}, \dots, \SI{28}{nm} \right\rbrace,
\end{equation}
i.e., we chose an increment of $\SI{2}{nm}$. For the particle anisotropy constant $K^\text{anis}$ we chose
\begin{equation}
    \mathcal{P}_\text{anis} = \left\lbrace \SI{100}{Jm^{-3}}, \SI{200}{Jm^{-3}}, \dots, \SI{9000}{Jm^{-3}} \right\rbrace,
\end{equation}
i.e., an increment of $\SI{100}{Jm^{-3}}$. Finally for the angles $\mathcal{P}_\text{angle}$ we took the same angles as used for the measurements.



\section{Results}

\subsection{Parameter Optimization}

\begin{figure}[t!]
    \centering
    \includegraphics[width=\columnwidth]{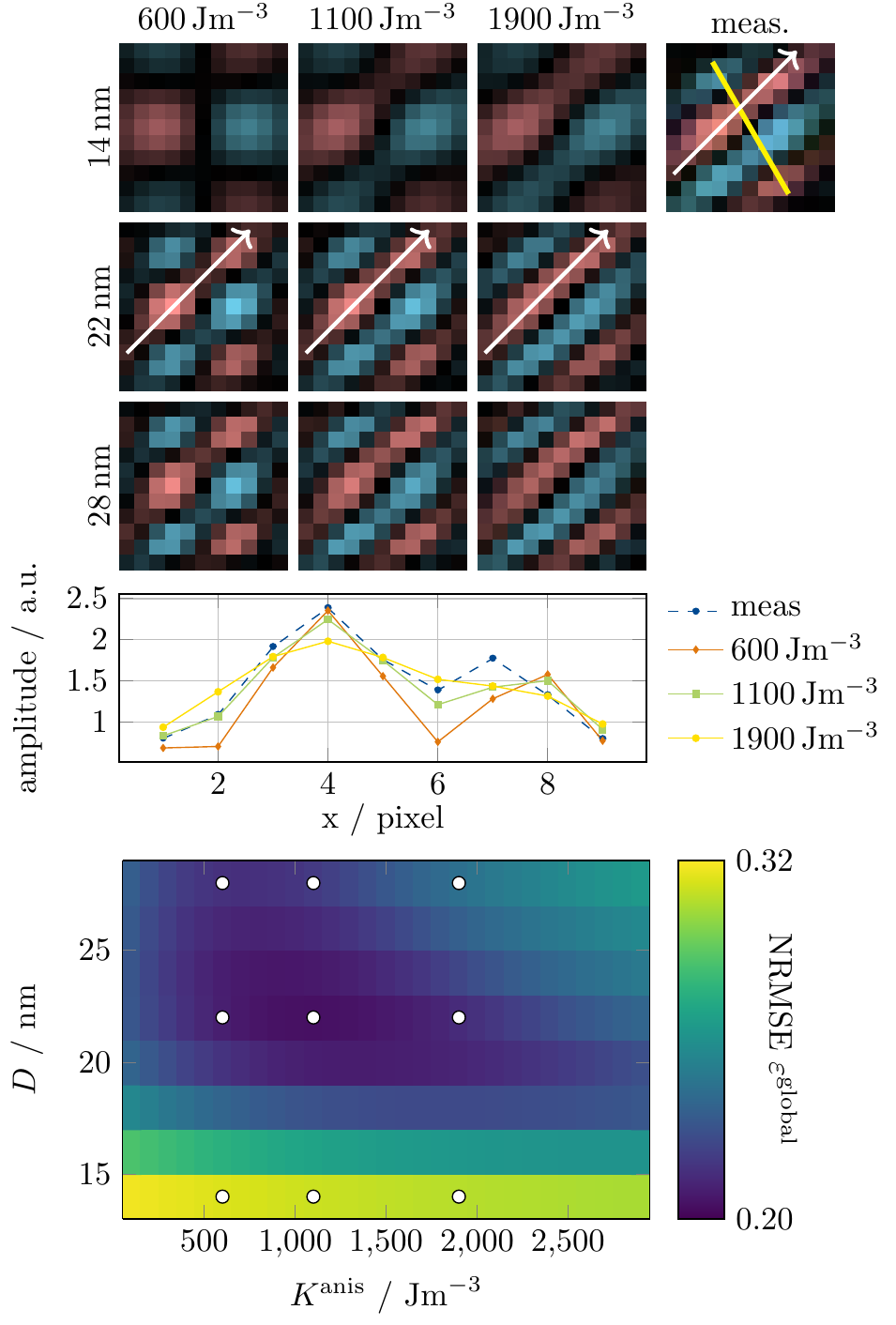}
    \caption{Parameter optimization for the model M1. The bottom shows the global NRMSE $\varepsilon^\text{global}$ combined for $x$- and $y$-channel as a function of the particle diameter $D$ and the particle anisotropy $K^\text{anis}$. The minimum of the discrete error map is reached at ($D=\SI{22}{nm}$, $K^\text{anis}=\SI{1100}{Jm^{-3}}$). For selected parameters indicated by white circles, the top part of the figure shows one row ($m_x=3$, $m_y=3$) of the modeled system matrices for the $x$-channel and the angle $\vartheta = 30^\circ$ in comparison with the measured system matrix shown in the last column. The yellow arrow indicates the easy axis for the considered angle. To highlight the influence of a change of the particle anisotropy, the central part of the figure shows a profile through the merging wave hills of the system matrix pattern. The profile is indicated as a white arrow in the considered system matrix pattern shown in the top part of the figure.}
    \label{fig:paramOptimization}
\end{figure}

The discrete parameter optimization for the equilibrium model M0 resulted in the diameter $D=\SI{22}{nm}$ as the optimal value minimizing \eqref{eq:globalMeasure}. For the more interesting case of M1, the optimization found the pair $(D,K^\text{anis})=(22\,\text{nm}, 1100\, \text{Jm}^{-3} )$ to be the global minimizer of \eqref{eq:globalMeasure}.

More details on the influence of the particle diameter and the anisotropy constant on the resulting system matrix are illustrated in Fig.~\ref{fig:paramOptimization} for a certain frequency component ($m_x=3$, $m_y=3$). In the top part of the figure the simulated system matrices and on the lower part the values of the error metric \eqref{eq:globalMeasure} are shown. First of all one can see that the error functional in the bottom part is a rather smooth curve without oscillations. Here, one can also see that the dimensions $D$ and $K^\text{anis}$ are not uncorrelated, which means that a larger particle diameter requires a smaller anisotropy constant to result in similar error values. Independent of these qualitative findings there is a minimum of the global error at $0.2049$. Selected system matrices shown above reveal the influence of the parameters $D$ and $K^\text{anis}$. Decreasing $D$ for fixed $K^\text{anis}$ leads to an increase of a spatial \textit{blurring} with respect to the offset field dimension, whereas this blurring has an isotropic nature. 
As the diameter influences the diffusion coefficient $\frac{1}{2\tau}$ in \eqref{eq:FP-general} only, when decreasing the diameter (resulting in larger $\frac{1}{2\tau}$), a larger offset field is necessary to obtain a similar effect. The blurring thus also includes a zooming effect to some degree. But due the delicate interplay of drive field, anisotropy, and offset field in the field vector $b$ in \eqref{eq:FP-general}, zooming is not the only effect which contribute to the blurring such that we stick to the term \textit{isotropic blurring} as it is caused by the isotropic diffusion coefficient $\frac{1}{2\tau}$ .  
In contrast,  increasing $K^\text{anis}$ for fixed $D$ leads to an \textit{anisotropic blurring}, which is oriented perpendicularly to the easy axis indicated as a yellow line. 
As $\tau$ is independent of $K^\text{anis}$, this is mainly caused by the structure of $b$ (see \eqref{eq:convection}) and it becomes apparent for large $K^\text{anis}$.
In the case that the offset field is parallel to the easy axis, the easy axis component and the offset field component of $b$ align such that we can observe the pattern along the easy axis in the field of view. The pattern in this direction is caused by the change of the offset field strength as it contributes effectively to the dominating anisotropy effect. 
If the offset field is perpendicular to the easy axis, an effective contribution by the offset field to the behavior of $b$ is small such that a larger offset field strength would be required to change the magnetization behavior significantly. This \textit{anisotropic blurring} effect can be seen in the top part of Fig.~\ref{fig:paramOptimization}, particularly in the right column.  
When comparing the simulated data with the measurement shown on the upper right, one can see that $(22\,\text{nm}, 1100 \,\text{Jm}^{-3} )$ yields the most similar looking pattern, which, as we discussed before, is also the value chosen based on the error metric.

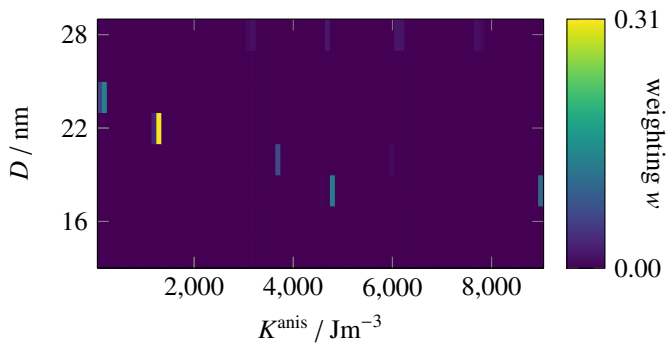
\begin{figure}
    \centering
    \begin{tikzpicture}[node distance=3pt and 3pt]

    \def\h{0.18\textwidth};
    \def\w{0.32\textwidth};
    
    \pgfplotsset{compat=1.6}
    \pgfplotsset{
    	matrixaxis/.style={
        	scale only axis,
            view={0}{90},
            enlargelimits=false,
            axis on top,
    		colormap/viridis,
            colorbar style={
    	        yticklabel style={
	                /pgf/number format/.cd,
	                fixed,
	                precision=2,
	                fixed zerofill,
    	        },
    	        ylabel={weighting $w$},
    	        ylabel style={
    	            yshift=5mm,
    	            rotate=180,
    	        },
            },
        },
    }
    
    \begin{axis}[matrixaxis, 
    			name=Weights,
    			width=\w, 
    			height=\h,
                point meta min=0.0,
                point meta max=0.31,
                ytick={16,22,28},
    			colorbar,
    			colorbar style={ytick={0.0,0.31}}, 
    			xlabel={$K^\textup{anis}$ / $\si{Jm}^{-3}$}, 
    			ylabel={$D$ / $\si{nm}$},
    			xlabel style={yshift=1pt},
    			ylabel style={yshift=4pt}]
    			
    	\addplot3[matrix plot*, mesh/cols=90] 
    	    table [x=K,y=D,z=weights, col sep=comma] 
    		{figures/img/weightmap.csv};
     
    \end{axis}
\end{tikzpicture}
\vspace{-0.7cm}
    \caption{Calculated weights $w_{D,K^\text{anis}}$ for model M2 determined by using the dictionary summarized in \eqref{eq:LSfitM2}.}
    \label{fig:weights}
\end{figure}

Considering M2, the TSVD-regularized (see Section~\ref{sec:data_driven_models}) optimization was performed for different regularization parameters $s$. Here, the regularization parameter takes the form of the number of singular values that were taken into account, i.e., the rank of the low-rank approximation of $M$. Lower numbers correspond to more aggressive rank reduction, and thus to stronger regularization. The values of the NRMSE are displayed in Table~\ref{table:svd}. We observe that the rank can be reduced quite aggressively before the results start to deteriorate. Only if less than 3\% of the largest singular values are considered, the error starts growing significantly. For that reason, we chose this value as the best balance between accuracy and interpretability. In other words, the 720 parameter combinations considered can be reduced to 21 effective structures by removing redundant information, resulting in a model more robust to measurement inaccuracies and noise.
The obtained weight vector $w$ for M2 is illustrated in Fig.~\ref{fig:weights}. 
Here we can see  similarities to the optimal parameter of model M1 as this $(D,K^\text{anis})$ tuple got a large weight value. But we can also observe smaller contributions by surrounding particle diameters and different anisotropy values. For instance, the two smallest simulated anisotropy constants $K^\text{anis} = \text{\SIlist{100;200}{Jm^{-3}}}$ are assigned nonzero weights for $D = \SI{24}{nm}$. The weights for different anisotropy constants may be interpreted in part as a hint towards the shape distribution of the particles, since shape anisotropy plays a large role in the overall anisotropic behavior~\cite{Cullity2008a}. In this context, small anisotropies hint towards more spherical particles while larger anisotropies may stand for ellipsoid shapes. There are also small contributions by the largest simulated diameter for various anisotropy constants. This may reflect larger particles or even clusters. In general, the role of clusters and chains in this consideration is unclear. We expect that the polydisperse nature as well as agglomeration effects lead to the observed structures. Consequently, the weights should serve as a hint at best for the true distribution of nanoparticle properties.

\begin{table}
    \centering
    \begin{tabular}{c|c}
     \% of SVs (Matrix rank $s$)  & NRMSE $\varepsilon^\text{global}$ \\ \hline
     100 (720) & 0.1994\\
     5 (36) & 0.1995\\
     4  (28) & 0.1995\\
     \bf{3} (\bf{21}) & \bf{0.1996}\\
     2  (14) & 0.2005\\
     1  (7) & 0.2034\\
     0.5 (3) & 0.2178 \\ 
    \end{tabular}
    \caption{Global NRMSE $\varepsilon^\text{global}$ for the truncated SVD approach for different numbers of singular values considered and the corresponding matrix ranks.}
    \label{table:svd}
\end{table}

\subsection{Qualitative Comparison}

\begin{figure*}[t!]
    \centering
    \includegraphics[width=0.99\textwidth]{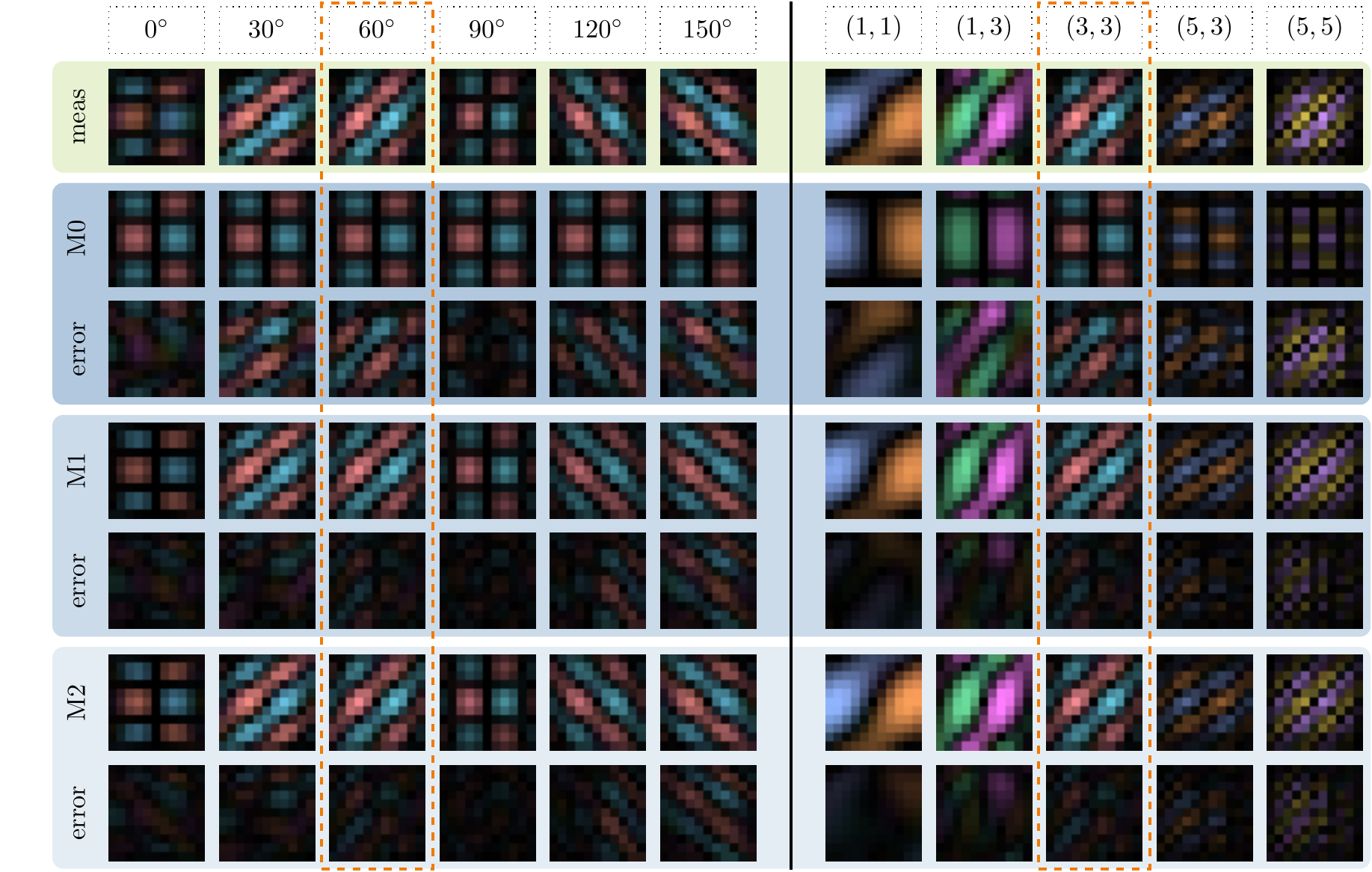}
    \caption{Comparison of the modeled MPI system matrices for models M0 to M2 with the measured system matrix for the $x$-channel. The left part of the figure shows a comparison for a fixed frequency component ($m_x=3$, $m_y=3$) and various angles in $30^\circ$ steps. The right part of the figure shows the data for a fixed angle of $60^\circ$ and several frequency components at mixing orders $(m_x, m_y) \in \{(1,1),(1,3),(3,3),(5,3),(5,5)\}$. The orange dashed box indicates the data that is present in both parts of the plot. In addition to the matrix patterns itself, the complex-valued difference $D_{k,\vartheta} = S^\text{model}_{k,\vartheta} - S^\text{calib}_{k,\vartheta}$ is shown. All data uses the complex colormap shown in Fig.~\ref{fig1:intro_example} whereas the same windowing is used for each frequency index $k$.}
    \label{fig:SF1}
\end{figure*}

We proceed with a qualitative comparison of the optimized models, M0 to M2. Fig.~\ref{fig:SF1} shows parts of the modeled and measured system matrices. Since the matrices are too large such that one could show them at once, we show selected parts sampled along different axes. On the left part of the figure the frequency component for $m_x=3$, $m_y=3$ is again selected but this time every second measured angle is shown. Here one can see that model M0 is not capable of modeling the effect of the anisotropic and angle dependent blurring and just provides a mean model with no merge of wave hills and no intensity variations for the different angles. In contrast, M1 much better resembles the measurement for the chosen frequency component and one can clearly see both features: An angle dependent anisotropic blurring and an angle dependent change in intensity. This can also be seen in the different frequency components for a fixed angle, which are shown on the right hand side of the figure. To highlight the differences between the measurement and the model, Fig.~\ref{fig:SF1} also shows difference maps that highlight the angle dependent deviation for M0 and the much lower overall deviation for the models M1 and M2. When switching to the model M2, differences to model M1 are less pronounced. For $(m_x=1,m_y=3)$ one can see that the first pixel row has a sightly higher value for M2, which better resembles the intensity for the measured frequency component.



\subsection{Quantitative Comparison}

\begin{figure*}[t!]
    \centering
    \includegraphics[width=0.99\textwidth]{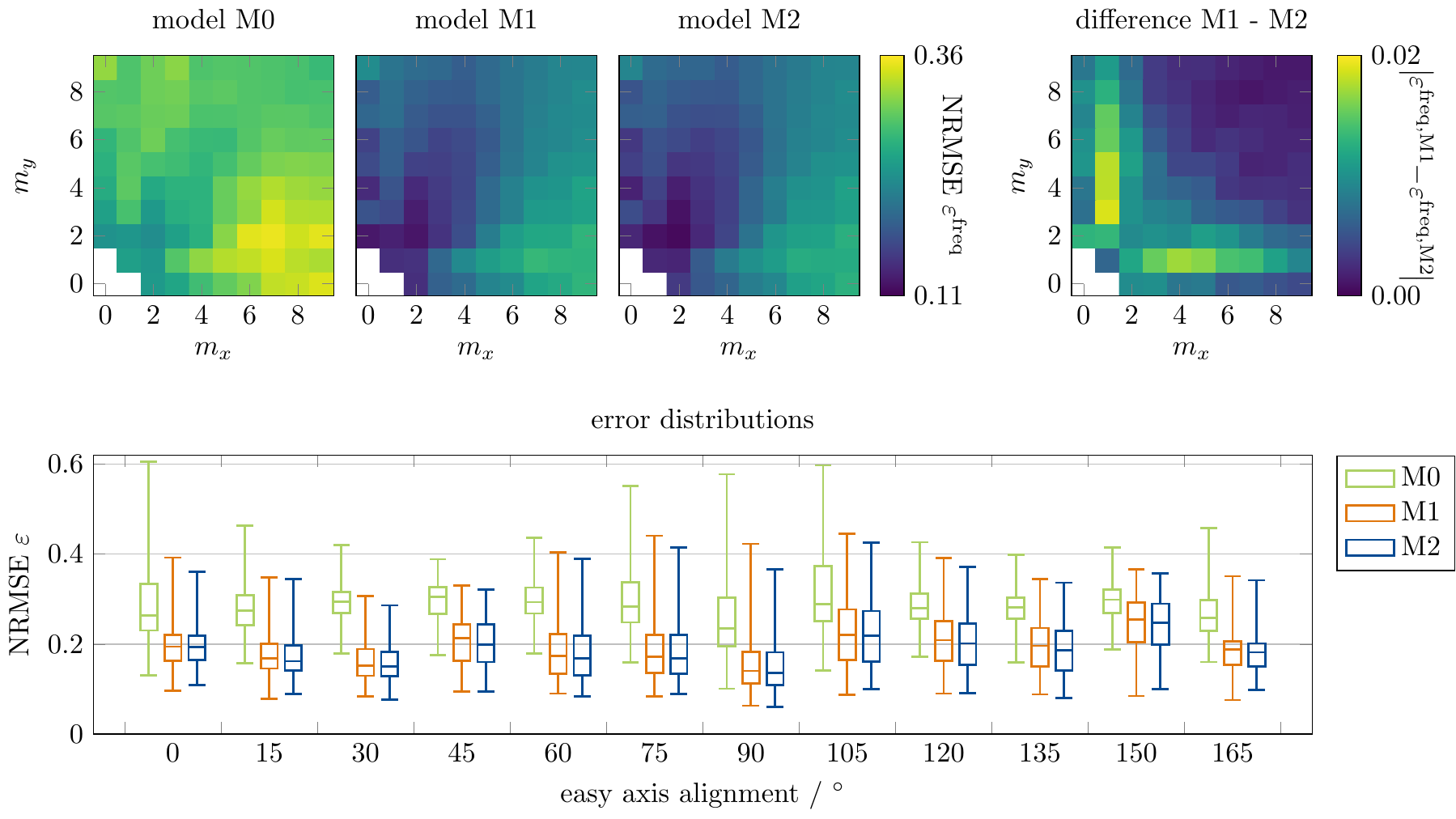}
    \caption{Differences between the measured and modeled system matrices for models M0--M2. On the top the error $\varepsilon^\text{freq}_{k}$ is illustrated for mixing orders $m_x = 0,\dots, 9$ and  $m_y = 0,\dots, 9$, while the individual errors along the angle and channel dimensions have been combined. The values for $(m_x,m_y) \in \{ (0,0), (0,1),(1,0) \}$ have been omitted since they do not contain any particle signal in the measurements (transfer function close to zero). The lower plot shows a statistical analysis of the error $\varepsilon$ as a box plot. Here, the statistical analysis has been performed along the frequency and channel dimensions, while for each angle an individual analysis has been performed.
    }
    \label{fig:errorMixOrders}
\end{figure*}

After the qualitative comparison of the modeled MPI system matrices with the measured one we next carry out a quantitative comparison. Here, we consider the distance measures introduced in Section~\ref{sec:distance_measure}. We first focus on the frequency dependent measure $\varepsilon_k^\text{freq}$ where the data for all angles and receive channels has been combined into a single number per frequency index $k$, respectively mixing order pair $(m_x,m_y)$. The resulting errors are shown in the upper part of Fig.~\ref{fig:errorMixOrders}. The first finding is that the models M1 and M2 result in considerably lower differences to the measurements than the equilibrium model, which confirms the results of the qualitative comparison. When comparing the errors for M1 and M2, one can hardly see a difference, which underlines the observation that we made in the qualitative comparison. To highlight that there are smaller differences we plot the difference $\left\vert\varepsilon^\text{freq,M1}_k- \varepsilon^\text{freq,M2}_k\right\vert$ in the upper right part of the figure. One can see that model M2 shows a slight improvement that is two orders of magnitude lower than the actual error. The improvements is highest when one of both mixing orders is 1.


Next we investigate the deviation in dependence of the considered easy axis angle. Here, we perform a statistical analysis on the differences $\varepsilon$ and set up a box-whisker plot for each angle and each model in the lower part of Fig.~\ref{fig:errorMixOrders}. First of all, one can see again that models M1 and M2 outperform the equilibrium model M0 and show much lower error. Interestingly, one can also see an angular dependency of the median error, which varies between $0.141$ and $0.255$ for model M1 and between $0.234$ and $0.305$ for model M0. Finally, the global NRMSE for the three considered models is summarized in Table~\ref{tab:globalError}, which wraps up our quantitative comparison.

\begin{table}
    \renewcommand{\arraystretch}{1.2}
    \centering
    \begin{tabular}{c|c|c|c}
         & M0 & M1 & M2 \\ \hline
     NRMSE $\varepsilon^\text{global}$ &  0.2976 & 0.2049 & 0.1996 \\
    \end{tabular}
    \caption{Global NRMSE $\varepsilon^\text{global}$ between the modeled system matrices for M0 to M2 and the measured system matrices.}
    \label{tab:globalError}
    \renewcommand{\arraystretch}{1.0}
\end{table}



\section{Discussion and Outlook}


Simulating the particle magnetization in magnetic particle imaging remains a very challenging task, which is due to the complex interplay of the particles on a micromagnetic level and a large possible parameter space involving size and anisotropy which is not exactly known for a set of more than billions of particles that is typically located in an image voxel and contributes a signal during a typical MPI experiment. One thus needs to consider ensembles which then only can describe the physical phenomenon in a mean sense.


In this work we progressed in the development of tools for describing MPI experiments based on physical models. Our approach uses in its core a magnetization model exploiting the Fokker-Plank equation, making the assumption that particle-particle interaction as well as agglomeration effects can be neglected. The physical model is then enriched by some data-driven components to bring the model in alignment with experimental data. First of all we still use a fitted transfer function, which simplifies model-based approaches tremendously since it allows to account for amplitude and phase differences that can be caused by an absent / inaccurate transfer function of the receive chain or by imperfectly estimated particle and scanner parameters. Secondly, we use a data-driven approach for obtaining the particle parameters (diameter and anisotropy constant). Here we used two different approaches. One simply pics the parameters that best describe the experiment while the other allows for linear combination of simulations obtained for different parameter sets. This dictionary approach is very flexible but it also makes physically non-plausible solutions possible if no physical priors are integrated, which we did not in this study. Our results indicate that the dictionary approach only slightly improves the result over the single parameter set simulation in our setting. 
A natural extension of M2 is a joint reconstruction of one transfer function for each receive channel only and the weight vector simultaneously. On the one hand this models the physical relationship more properly, but on the other hand its solution requires usage of more sophisticated nonlinear programming tools. This direction of research remains immediate future work. 


In our experiment we used an immobilized particle setup with oriented easy axis, which marks both an interesting and challenging setup for model studies since it allows to change the orientation as a parameter and to compare if the model can describe the changes in the measured signal. As expected, the equilibrium model can not describe changes in easy axis orientation, whereas both our models M1 and M2 were successful in describing the observed phenomenons which manifest themselves as an anisotropic blurring of the system matrix patterns in perpendicular direction to the magnetic easy axis.
Of course, this might also be an interplay between easy axis orientation and the Lissajous pattern.
Future works should include the study of different excitation patterns and might also include excitation optimization for multi-contrast reconstruction with respect to easy axis orientations. 
We note that there was some angular dependency in the model accuracy, which cannot be explained by the model itself. One reason might be that the positioning of the sample or the stability of the MPI scanner did slightly change leading to a higher deviation for, in particular, the $150^\circ$ case.


Our focus in this work was not on the deployment of a simulation-based model, which would require additional work. First of all we used a fully measured system matrix for parameter estimation (transfer function, particle diameter and particle anisotropy), which would not be available in practice since the modeled system matrix aims to replace the measurement. One route that has already been exploited for the equilibrium model in~\cite{Knopp2010d} is to rely on a low-resolution based system matrix, which has the advantage that its calibration scans can be done in few minutes and one still has reference data to compare the model with. This can be combined with compressed sensing based system matrix recovery~\cite{knopp2013sparse,grosser2020using} or simultaneous high resolution system matrix and concentration reconstruction \cite{KluthBathkeJiangMaass2020}. In particular, motivated by the structure of the reconstructed weight vector in this work, a sparsity-promoting approach with respect to the dictionary obtained by the physically-driven model we developed can be highly beneficial to reduce the amount of necessary calibration data. We also did not deploy the modeled system matrices for image reconstruction yet since this is its own topic and needs more sophisticated phantoms in case of immobilized samples with aligned easy axis. Subject of a future study will be the development of an immobilized phantom with differently oriented samples in the FOV such that the evaluation can be done based on the multi-contrast reconstruction outlined in~\cite{Moeddel2020IWMPI}.


We note that the considered model-system with axis-aligned easy axis is not just of theoretical interest but has clear application in medical imaging scenarios in particular in the field of interventional imaging where the instruments will be coated with immobilized particles that can be additionally orientated.
This would enable the determination of position and orientation of a medical instrument such as a catheter, guide wire, or stent. As discussed in~\cite{Moeddel2020IWMPI}, a large number of calibration measurements is necessary for image reconstruction in this scenario, which will require model-based approaches for system matrix calibration since the calibration time on a fine spatial grid would be prohibitive in case of 3D excitation, which would be employed in a clinical environment. Our work contributes to this goal and provides a platform for future research where the goal will be to remove the remaining minor but still observable model-measurement mismatch that our model still has.

%
%

\section*{Acknowledgements}

H. Albers and T. Kluth acknowledge funding by the German Research Foundation (DFG, Deutsche Forschungsgemeinschaft) - project 426078691.



 \bibliographystyle{elsarticle-num} 
 \bibliography{ref,mypub,cas-refs}

\begin{thebibliography}{10}
\expandafter\ifx\csname url\endcsname\relax
  \def\url#1{\texttt{#1}}\fi
\expandafter\ifx\csname urlprefix\endcsname\relax\def\urlprefix{URL }\fi
\expandafter\ifx\csname href\endcsname\relax
  \def\href#1#2{#2} \def\path#1{#1}\fi

\bibitem{krishnan2010biomedical}
K.~M. Krishnan, Biomedical nanomagnetics: a spin through possibilities in
  imaging, diagnostics, and therapy, IEEE transactions on magnetics 46~(7)
  (2010) 2523--2558.

\bibitem{le2017real}
T.-A. Le, X.~Zhang, A.~K. Hoshiar, J.~Yoon, Real-time two-dimensional magnetic
  particle imaging for electromagnetic navigation in targeted drug delivery,
  Sensors 17~(9) (2017) 2050.

\bibitem{zhang2017development}
X.~Zhang, T.-A. Le, J.~Yoon, Development of a real time imaging-based guidance
  system of magnetic nanoparticles for targeted drug delivery, Journal of
  Magnetism and Magnetic Materials 427 (2017) 345--351.

\bibitem{griese2020simultaneous}
F.~Griese, T.~Knopp, C.~Gruettner, F.~Thieben, K.~M{\"u}ller, S.~Loges,
  P.~Ludewig, N.~Gdaniec, Simultaneous magnetic particle imaging and navigation
  of large superparamagnetic nanoparticles in bifurcation flow experiments,
  Journal of Magnetism and Magnetic Materials 498 (2020) 166206.

\bibitem{knopp2017magnetic}
T.~Knopp, N.~Gdaniec, M.~M{\"o}ddel, Magnetic particle imaging: From proof of
  principle to preclinical applications, Phys. Med. Biol. 62~(14) (2017) R124.
\newblock \href {https://doi.org/10.1088/1361-6560/aa6c99}
  {\path{doi:10.1088/1361-6560/aa6c99}}.

\bibitem{herz2017magnetic}
S.~Herz, P.~Vogel, T.~Kampf, M.~R{\"u}ckert, S.~Veldhoen, V.~C. Behr, T.~A.
  Bley, Magnetic particle imaging for quantification of vascular stenoses: a
  phantom study, IEEE transactions on medical imaging 37~(1) (2017) 61--67.

\bibitem{kaul2018magnetic}
M.~G. Kaul, J.~Salamon, T.~Knopp, H.~Ittrich, G.~Adam, H.~Weller, C.~Jung,
  Magnetic particle imaging for in vivo blood flow velocity measurements in
  mice, Physics in Medicine \& Biology 63~(6) (2018) 064001.

\bibitem{vogel2020superspeed}
P.~Vogel, M.~A. R{\"u}ckert, T.~Kampf, S.~Herz, A.~Stang, L.~W{\"o}ckel, T.~A.
  Bley, S.~Dutz, V.~C. Behr, Superspeed bolus visualization for vascular
  magnetic particle imaging, IEEE transactions on medical imaging 39~(6) (2020)
  2133--2139.

\bibitem{zhu2019quantitative}
X.~Zhu, J.~Li, P.~Peng, N.~Hosseini~Nassab, B.~R. Smith, Quantitative drug
  release monitoring in tumors of living subjects by magnetic particle imaging
  nanocomposite, Nano letters 19~(10) (2019) 6725--6733.

\bibitem{khandhar2017evaluation}
A.~Khandhar, P.~Keselman, S.~Kemp, R.~Ferguson, P.~Goodwill, S.~Conolly,
  K.~Krishnan, Evaluation of peg-coated iron oxide nanoparticles as blood pool
  tracers for preclinical magnetic particle imaging, Nanoscale 9~(3) (2017)
  1299--1306.

\bibitem{kaul2017vitro}
M.~G. Kaul, T.~Mummert, C.~Jung, J.~Salamon, A.~P. Khandhar, R.~M. Ferguson,
  S.~J. Kemp, H.~Ittrich, K.~M. Krishnan, G.~Adam, et~al., In vitro and in vivo
  comparison of a tailored magnetic particle imaging blood pool tracer with
  resovist, Phys. Med. Biol. 62~(9) (2017) 3454.
\newblock \href {https://doi.org/10.1088/1361-6560/aa5780}
  {\path{doi:10.1088/1361-6560/aa5780}}.

\bibitem{salamon2016magnetic}
J.~Salamon, M.~Hofmann, C.~Jung, M.~Kaul, F.~Werner, K.~Them, R.~Reimer,
  P.~Nielsen, A.~vom Scheidt, G.~Adam, et~al., Magnetic particle/magnetic
  resonance imaging: In-vitro {MPI}-guided real time catheter tracking and {4D}
  angioplasty using a road map and blood pool tracer approach, PloS one 11~(6)
  (2016) e0156899.
\newblock \href {https://doi.org/10.1371/journal.pone.0156899}
  {\path{doi:10.1371/journal.pone.0156899}}.

\bibitem{rahmer2017interactive}
J.~Rahmer, D.~Wirtz, C.~Bontus, J.~Borgert, B.~Gleich, Interactive magnetic
  catheter steering with {3-D} real-time feedback using multi-color magnetic
  particle imaging, IEEE Trans. Med. Imaging 36~(7) (2017) 1449--1456.
\newblock \href {https://doi.org/10.1109/TMI.2017.2679099}
  {\path{doi:10.1109/TMI.2017.2679099}}.

\bibitem{herz2019magnetic}
S.~Herz, P.~Vogel, T.~Kampf, P.~Dietrich, S.~Veldhoen, M.~A. R{\"u}ckert,
  R.~Kickuth, V.~C. Behr, T.~A. Bley, Magnetic particle imaging--guided
  stenting, Journal of Endovascular Therapy 26~(4) (2019) 512--519.

\bibitem{Moeddel2020IWMPI}
M.~M{\"o}ddel, F.~Griese, T.~Kluth, T.~Knopp, Estimating orientation using
  multi-contrast {MPI}, International Journal on Magnetic Particle Imaging 6~(2
  Suppl. 1) (2020) ID 2009023, 3 pages.

\bibitem{rahmer2015first}
J.~Rahmer, A.~Halkola, B.~Gleich, I.~Schmale, J.~Borgert, First experimental
  evidence of the feasibility of multi-color magnetic particle imaging, Phys.
  Med. Biol. 60~(5) (2015) 1775--91.

\bibitem{MoeddelGrieseKluthKnopp2021_preprint}
M.~M\"oddel, F.~Griese, T.~Kluth, T.~Knopp, Spatial orientation estimation of
  immobilized magnetic nanoparticles with parallel aligned easy axes, Preprint
  (2021).

\bibitem{Knopp2010PhysMedBio}
T.~Knopp, J.~Rahmer, T.~Sattel, S.~Biederer, J.~Weizenecker, B.~Gleich,
  J.~Borgert, T.~Buzug, Weighted iterative reconstruction for magnetic particle
  imaging, Physics in Medicine and Biology 55~(6) (2010) 1577 -- 1589.
\newblock \href {https://doi.org/10.1088/0031-9155/55/6/003}
  {\path{doi:10.1088/0031-9155/55/6/003}}.

\bibitem{grosser2020using}
M.~Grosser, M.~M{\"o}ddel, T.~Knopp, Using low-rank tensors for the recovery of
  {MPI} system matrices, IEEE Transactions on Computational Imaging 6 (2020)
  1389--1402.

\bibitem{knopp2009model}
T.~Knopp, T.~F. Sattel, S.~Biederer, J.~Rahmer, J.~Weizenecker, B.~Gleich,
  J.~Borgert, T.~M. Buzug, Model-based reconstruction for magnetic particle
  imaging, IEEE Transactions on Medical Imaging 29~(1) (2009) 12--18.

\bibitem{knopp20102d}
T.~Knopp, S.~Biederer, T.~F. Sattel, J.~Rahmer, J.~Weizenecker, B.~Gleich,
  J.~Borgert, T.~M. Buzug, {2D} model-based reconstruction for magnetic
  particle imaging, Medical Physics 37~(2) (2010) 485--491.

\bibitem{Kluth2018a}
T.~Kluth, Mathematical models for magnetic particle imaging, Inverse Problems
  34~(8) (2018) 083001.

\bibitem{KluthSzwargulskiKnopp2019}
T.~Kluth, P.~Szwargulski, T.~Knopp, Towards accurate modeling of the
  multidimensional magnetic particle imaging physics, New Journal of Physics
  21~(10) (2019) 103032.
\newblock \href {https://doi.org/10.1088/1367-2630/ab4938}
  {\path{doi:10.1088/1367-2630/ab4938}}.

\bibitem{Coffey1992}
W.~T. Coffey, P.~J. Cregg, Y.~U.~P. Kalmykov, On the Theory of Debye and
  N\'{e}el Relaxation of Single Domain Ferromagnetic Particles, John Wiley \&
  Sons, Inc., 1992, pp. 263--464.
\newblock \href {https://doi.org/10.1002/9780470141410.ch5}
  {\path{doi:10.1002/9780470141410.ch5}}.

\bibitem{shliomis1994theory}
M.~Shliomis, V.~Stepanov, Theory of the dynamic susceptibility of magnetic
  fluids, Advances in Chemical Physics: Relaxation Phenomena in Condensed
  Matter 87 (1994) 1--30.

\bibitem{Weizenecker2010particle}
J.~Weizenecker, B.~Gleich, J.~Rahmer, J.~Borgert, Particle dynamics of
  mono-domain particles in magnetic particle imaging, in: Magnetic
  Nanoparticles, World Scientific, 2010, pp. 3--15.
\newblock \href {https://doi.org/10.1142/9789814324687\_0001}
  {\path{doi:10.1142/9789814324687\_0001}}.

\bibitem{Yoshida2012}
T.~Yoshida, K.~Enpuku, Nonlinear behavior of magnetic fluid in brownian
  relaxation: Numerical simulation and derivation of empirical model, in: T.~M.
  Buzug, J.~Borgert (Eds.), Magnetic Particle Imaging, Springer Berlin
  Heidelberg, Berlin, Heidelberg, 2012, pp. 9--13.

\bibitem{weizenecker2012micro}
J.~Weizenecker, B.~Gleich, J.~Rahmer, J.~Borgert, Micro-magnetic simulation
  study on the magnetic particle imaging performance of anisotropic mono-domain
  particles, Phys. Med. Biol. 57~(22) (2012) 7317.

\bibitem{Yoshida2012b}
T.~Yoshida, K.~Enpuku, J.~Dieckhoff, M.~Schilling, F.~Ludwig, Magnetic fluid
  dynamics in a rotating magnetic field, J. Appl. Phys. 111~(5) (2012) 053901.

\bibitem{yoshida2013characterization}
T.~Yoshida, N.~Othman, K.~Enpuku, Characterization of magnetically fractionated
  magnetic nanoparticles for magnetic particle imaging, J. Appl. Phys. 114~(17)
  (2013) 173908.

\bibitem{rogge2013simulation}
H.~Rogge, M.~Erbe, T.~M. Buzug, K.~L{\"u}dtke-Buzug, Simulation of the
  magnetization dynamics of diluted ferrofluids in medical applications,
  Biomed. Tech. 58~(6) (2013) 601--609.

\bibitem{reeves2014approaches}
D.~B. Reeves, J.~B. Weaver, Approaches for modeling magnetic nanoparticle
  dynamics, Crit. Rev. Biomed. Eng. 42~(1) (2014) 85--93.

\bibitem{Martens2013}
M.~Martens, R.~Deissler, Y.~Wu, L.~Bauer, Z.~Yao, R.~Brown, M.~Griswold,
  Modeling the {B}rownian relaxation of nanoparticle ferrofluids: Comparison
  with experiment, Med. Phys. 40~(2) (2013) 022303.

\bibitem{Deissler2014}
R.~J. Deissler, Y.~Wu, M.~A. Martens, Dependence of {B}rownian and {N}\'{e}el
  relaxation times on magnetic field strength, Med. Phys. 41~(1) (2014) 012301,
  1--12.

\bibitem{Enpuku2014}
K.~Enpuku, S.~Bai, A.~Hirokawa, K.~Tanabe, T.~Sasayama, T.~Yoshida, The effect
  of neel relaxation on the properties of the third harmonic signal of magnetic
  nanoparticles for use in narrow-band magnetic nanoparticle imaging, Jpn. J.
  Appl. Phys. 53~(10) (2014) 103002.

\bibitem{Shah2015}
S.~A. Shah, D.~B. Reeves, R.~M. Ferguson, J.~B. Weaver, K.~M. Krishnan, Mixed
  brownian alignment and n\'eel rotations in superparamagnetic iron oxide
  nanoparticle suspensions driven by an ac field, Phys. Rev. B 92 (2015)
  094438.

\bibitem{graeser2015trajectory}
M.~Graeser, K.~Bente, A.~Neumann, T.~M. Buzug, Trajectory dependent particle
  response for anisotropic mono domain particles in magnetic particle imaging,
  J. Phys. D: Appl. Phys. 49~(4) (2016) 045007.

\bibitem{Yoshida2017}
T.~Yoshida, Y.~Matsugi, N.~Tsujimura, T.~Sasayama, K.~Enpuku, T.~Viereck,
  M.~Schilling, F.~Ludwig, Effect of alignment of easy axes on dynamic
  magnetization of immobilized magnetic nanoparticles, J. Magn. Magn. Mater.
  427 (2017) 162 -- 167.

\bibitem{elrefai2021effect}
A.~L. Elrefai, K.~Enpuku, T.~Yoshida, Effect of easy axis alignment on dynamic
  magnetization of immobilized and suspended magnetic nanoparticles, Journal of
  Applied Physics 129~(9) (2021) 093905.

\bibitem{maass2020representation}
M.~Maass, A.~Mertins, On the representation of magnetic particle imaging in
  fourier space, International Journal on Magnetic Particle Imaging 6~(1)
  (2020).

\bibitem{von2017hybrid}
A.~von Gladiss, M.~Graeser, P.~Szwargulski, T.~Knopp, T.~Buzug, Hybrid system
  calibration for multidimensional magnetic particle imaging, Phys. Med. Biol.
  62~(9) (2017) 3392.

\bibitem{Kluth2019numerical}
T.~Kluth, B.~Jin, Enhanced reconstruction in magnetic particle imaging by
  whitening and randomized {SVD} approximation, Phys. Med. Biol. 64~(12) (2019)
  125026.

\bibitem{AlbersKluthKnopp2020_preprint}
H.~Albers, T.~Kluth, T.~Knopp, A simulation framework for particle
  magnetization dynamics of large ensembles of single domain particles:
  Numerical treatment of {B}rown/{N}\'{e}el dynamics and parameter
  identification problems in magnetic particle imaging., Preprint, arXiv:
  2010.07772 (2020).

\bibitem{Kluth2017}
T.~Kluth, P.~Maass, Model uncertainty in magnetic particle imaging: Nonlinear
  problem formulation and model-based sparse reconstruction, International
  Journal on Magnetic Particle Imaging 3~(2) (2017) ID 1707004, 10 pages.

\bibitem{optimization}
P.~E. Gill, W.~Murray, M.~H. Wright, Practical Optimization, Society for
  Industrial and Applied Mathematics, Philadelphia, PA, 2019.
\newblock \href {https://doi.org/10.1137/1.9781611975604}
  {\path{doi:10.1137/1.9781611975604}}.

\bibitem{rahmer2012analysis}
J.~Rahmer, J.~Weizenecker, B.~Gleich, J.~Borgert, Analysis of a {3-D} system
  function measured for magnetic particle imaging, IEEE Trans. Med. Imaging
  31~(6) (2012) 1289--1299.

\bibitem{weber2015artifact}
A.~Weber, F.~Werner, J.~Weizenecker, T.~M. Buzug, T.~Knopp, Artifact free
  reconstruction with the system matrix approach by overscanning the
  field-free-point trajectory in magnetic particle imaging, Physics in Medicine
  \& Biology 61~(2) (2015) 475.

\bibitem{yoshida2017effect}
T.~Yoshida, Y.~Matsugi, N.~Tsujimura, T.~Sasayama, K.~Enpuku, T.~Viereck,
  M.~Schilling, F.~Ludwig, Effect of alignment of easy axes on dynamic
  magnetization of immobilized magnetic nanoparticles, Journal of Magnetism and
  Magnetic Materials 427 (2017) 162--167.

\bibitem{Elrefai2020}
A.~L. {Elrefai}, T.~{Sasayama}, T.~{Yoshida}, K.~{Enpuku}, Ac magnetization of
  immobilized magnetic nanoparticles with different degrees of parallel
  alignment of easy axes, IEEE Transactions on Magnetics (2020) 1--1\href
  {https://doi.org/10.1109/TMAG.2020.3024864}
  {\path{doi:10.1109/TMAG.2020.3024864}}.

\bibitem{Cullity2008a}
B.~D. Cullity, C.~D. Graham, Introduction to Magnetic Materials, Wiley-IEEE
  Press, 2008.

\bibitem{Knopp2010d}
T.~Knopp, S.~Biederer, T.~F. Sattel, J.~Rahmer, J.~Weizenecker, B.~Gleich,
  J.~Borgert, T.~M. Buzug, {2D} model-based reconstruction for magnetic
  particle imaging, Med. Phys. 37~(2) (2010) 485--491.

\bibitem{knopp2013sparse}
T.~Knopp, A.~Weber, Sparse reconstruction of the magnetic particle imaging
  system matrix, IEEE Trans. Med. Imaging 32~(8) (2013) 1473--1480.

\bibitem{KluthBathkeJiangMaass2020}
T.~Kluth, C.~Bathke, M.~Jiang, P.~Maass, Joint super-resolution image
  reconstruction and parameter identification in imaging operator: analysis of
  bilinear operator equations, numerical solution, and application to magnetic
  particle imaging, Inverse Problems 36~(12) (2020) 124006.
\newblock \href {https://doi.org/10.1088/1361-6420/abc2fe}
  {\path{doi:10.1088/1361-6420/abc2fe}}.

\end{thebibliography}





\end{document}